\documentclass[fleqn,usenatbib]{mnras}

\usepackage{newtxtext,newtxmath}
\usepackage[T1]{fontenc}

\DeclareRobustCommand{\VAN}[3]{#2}
\let\VANthebibliography\thebibliography
\def\thebibliography{\DeclareRobustCommand{\VAN}[3]{##3}\VANthebibliography}

\usepackage{graphicx}	% Including figure files
\usepackage{amsmath}	% Advanced maths commands
\title[Diagnosing of a filament-cavity flux rope system]{Stereoscopic diagnosing of a filament-cavity flux rope system by tracing the path of a two-sided-loop jet}

\author[Tan et al.]{
Song Tan,$^{1,3}$
Yuandeng Shen,$^{1,2,3,4}$\thanks{E-mail: ydshen@ynao.ac.cn}
Xinping Zhou,$^{1,3}$
Yadan Duan,$^{1,3}$
Zehao Tang,$^{1}$
Chengrui Zhou,$^{1,3}$
\newauthor
and Surui Yao$^{1}$
\\
\\
$^{1}$Yunnan Observatories, Chinese Academy of Sciences,  Kunming, 650216, China\\
$^{2}$State Key Laboratory of Space Weather, Chinese Academy of Sciences, Beijing 100190, China\\
$^{3}$University of Chinese Academy of Sciences, Beijing 100049, China\\
$^{4}$Yunnan Key Laboratory of Solar Physics and Space Exploration, Kunming, 650216, China 
}

\date{Accepted 2022-Jun-27. Received YYY; in original form ZZZ}

\pubyear{2022}

\begin{document}
\label{firstpage}
\pagerange{\pageref{firstpage}--\pageref{lastpage}}
\maketitle

\begin{abstract}
The fine magnetic structure is vitally important to understanding the formation, stabilization and eruption of solar filaments, but so far, it is still an open question yet to be resolved. Using stereoscopic observations taken by the Solar Dynamics Observatory and Solar TErrestrial RElations Obsevatory, we studied the generation mechanism of a two-sided-loop jet (TJ) and the ejection process of the jet plasma into the overlying filament-cavity system. We find that the generation of the two-sided-loop jet was due to the magnetic reconnection between an emerging flux loop and the overlying filament. The jet's two arms ejected along the filament axis during the initial stage. Then, the north arm bifurcated into two parts at about 50 Mm from the reconnection site. After the bifurcation, the two bifurcated parts were along the filament axis and the cavity which hosted the filament, respectively. By tracing the ejecting plasma flows of the TJ inside the filament, we not only
measured that the magnetic twist stored in the filament was at least 5$\pi$ but also found that the fine magnetic structure of the filament-cavity flux rope system is in well agreement with the theoretical results of Magnetic flux rope models.
\end{abstract}


%%%%%%%%%%%%%%%%%%%%%%%%%%%%%%%%%%%%%%%%%%%%%%%%%%

%%%%%%%%%%%%%%%%% BODY OF PAPER %%%%%%%%%%%%%%%%%%

\section{Introduction}

Magnetic flux ropes (MFRs) represent magnetic structures composed of wrapped magnetic fluxes around a central axis; they are believed to be the core  magnetic structures of many solar eruptions such as coronal mass ejections (CMEs). However, the basic fine magnetic structure of MFRs is still under hot debate \citep{2020RAA....20..165L}, although many observational and theoretical studies have been documented in the literature \citep{2011LRSP....8....1C, 2014LRSP...11....1P, 2018LRSP...15....7G}. Generally, MFR models explain the filament and cavity as two parts of the same magnetic structure \citep{2006JGRA..11112103G}. In H$\alpha$ and extreme ultraviolet observations of filament eruption, the rotation motion of filaments are often detected as the evidence of MFRs \citep{2003ApJ...595L.135J,2019ApJ...877L..28Z}; especially, the fine structure of filaments can be illuminated by hot plasma flows when solar jets inject into them \citep[e.g.,][]{2013ApJ...770L..25L, 2014ApJ...784L..36Y, 2019ApJ...883..104S}. However, the theoretical prediction that a MFR is composed of a filament and a cavity has not yet been supported by clear on-disk observations, although some limb observations have imaged distinct cavity hosting filaments or prominences \citep[e.g.,][]{2013ApJ...770L..28B, 2019ApJ...883..104S}. Since on-disk filaments show as dark absorption features, it is hard to diagnose their fine magnetic structures under normal conditions. Fortunately, this defect can be remedied by tracing mass flows along filaments or prominences \citep[e.g.,][]{2014ApJ...784L..36Y, 2015ApJ...814L..17S, 2021A&A...647A.112Z}. 

\begin{figure}
\centering
\includegraphics[width=8cm]{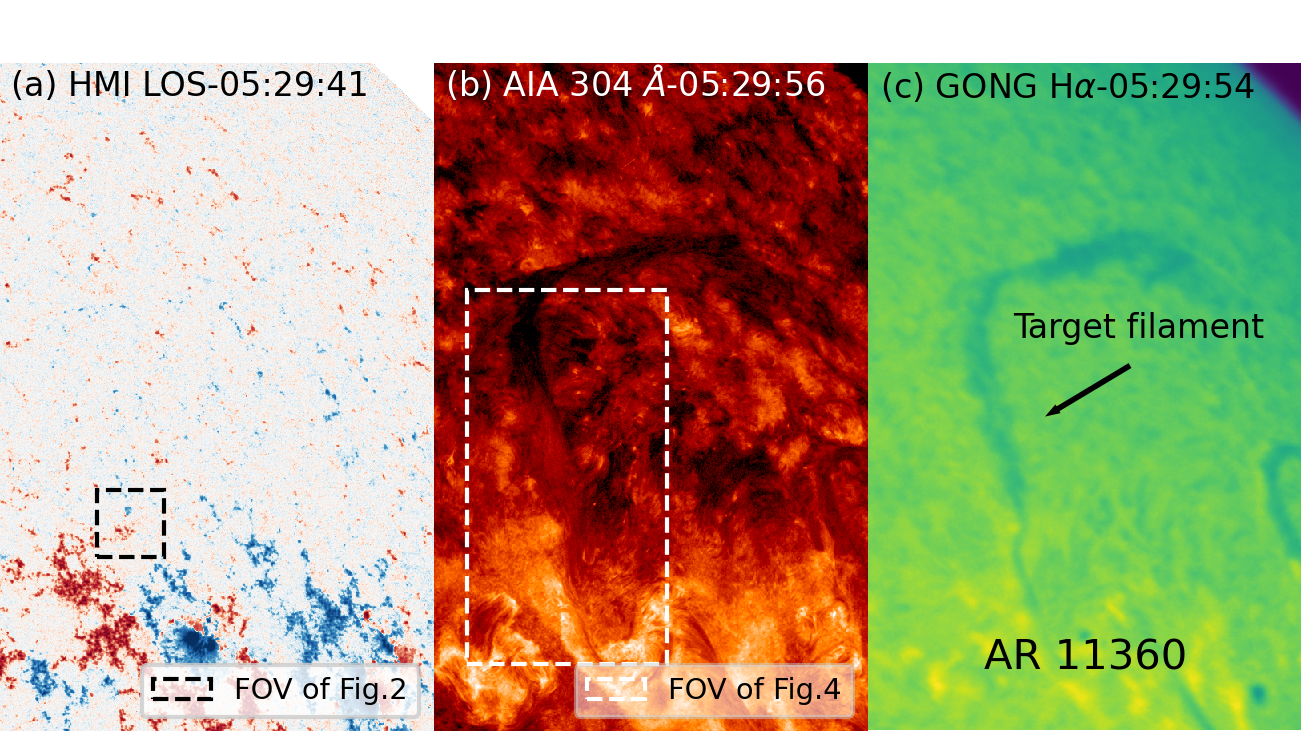}
\caption{(a) HMI LOS magnetogram covering the eruption source region and the filament. (b) The target filament in the AIA 304 \AA\ image at 05:30 UT. (C) The black arrow points to the filament in the GONG H$\alpha$ image. The polarity of the southern and northern foot points of the filament are negative and positive respectively.
}
\label{1}
\end{figure}

\begin{figure*}
\centering
\includegraphics[width=17cm]{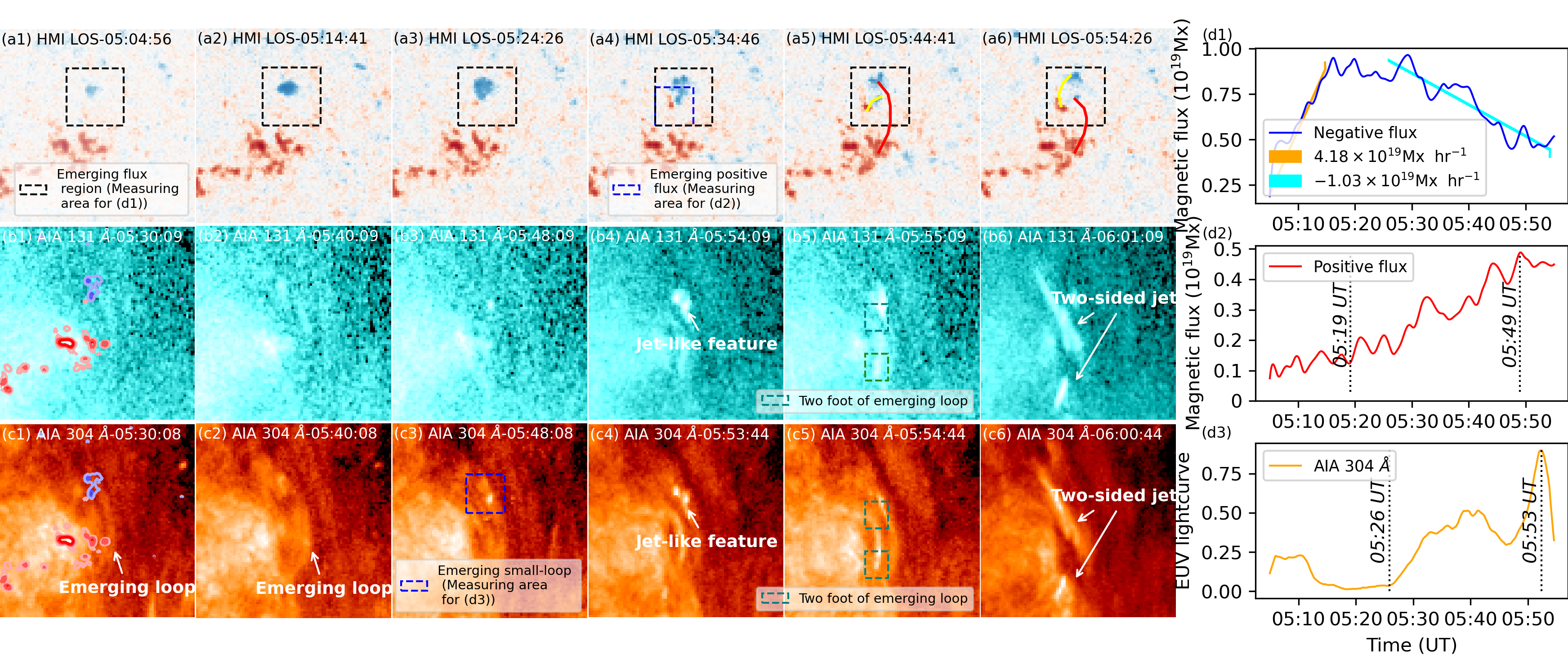}
\caption{Generation and evolution details of emerging flux loop in HMI LOS magnetograms and AIA (131 and 304 \AA)\ images superposed with HMI LOS magnetogram. The black dashed boxes in (a1)-(a6) indicate the emerging negative polarity of large flux loop. The blue dashed boxes in (a4), (b3) and (c3) show the appearance of small biploe positive polarity and emerging small loop, respectively. (d1) The plots of the negative polarity flux (absolute value) curve of large flux loop markedly with the growth and decreasing rate. (d2) and (d3) are the positive polarity flux curve and the AIA 304 \AA\ lightcurve of emerging small-loop, respectively. (An animation is provided online.)
}
\label{2}
\end{figure*}

A two-sided-loop jet (TJ) represents bi-directional plasma flows originating from the eruption source region \citep{2021RSPSA.47700217S}, which is generally thought to be caused by the magnetic reconnection between an emerging flux loop and the background horizontal field \citep[e.g.,][]{1995Natur.375...42Y,2013ApJ...775..132J,2018ApJ...861..108Z}, or between adjacent filament threads \citep{2017ApJ...845...94T}. However, recent high spatiotemporal observations showed that the generation of TJs is basically due to the eruptions of mini-filaments below filament magnetic systems \citep[e.g.,][]{2019ApJ...883..104S,2019ApJ...887..220Y,2020MNRAS.498L.104W} or surrounding horizontal magnetic field lines \citep[e.g.,][]{2019ApJ...871..220S}, resembling typical straight solar jets \citep[e.g.,][]{2012ApJ...745..164S,2017ApJ...851...67S,2015Natur.523..437S,2021ApJ...912L..15T}. \cite{2017ApJ...842...27Z} found that flare-generated jets moving along large-scale magnetic structures interact with remote filament, generating filament oscillations. Cornal jets can also trigger large-scale coronal waves \citep[][]{2018ApJ...861..105S,2018MNRAS.480L..63S,2018ApJ...860L...8S} and coronal mass ejections \citep[e.g.,][]{2008ApJ...677..699J}. Especially, solar jets occurred in filament channels are found to be of important to lead the oscillation, splitting, and even eruption of the overlying filaments \citep[e.g.,][]{2018NewA...65....7T}.

In this letter, we report the observations of a TJ below a large filament, whose material travels not only along the filament axis but also the hosting cavity. We investigate the generation mechanism of the TJ, and diagnose the fine magnetic structure of the overlying filament by tracing the ejecting jet mass flows with the aid of 3D reconstruction using stereoscopic observations taken by the Atmospheric Imaging Assembly \citep[AIA;][]{2012SoPh..275...17L} onboard the Solar Dynamics Observatory (SDO) and the Extreme Ultraviolet Imager \citep[EUVI;][]{2004SPIE.5171..111W} onboard the Solar TErrestrial RElations Observatory (STEREO). In addition, H$\alpha$ images from the Global Oscillation Network Group \citep[GONG;][]{1996Sci...272.1284H} and line-of-sight (LOS) magnetograms taken by the Helioseismic and Magnetic Imager (HMI) onboard the SDO are also used.

\section{Results}
\label{sec2}

\subsection{The generation of the Two-sided-loop jet}
The TJ originated in the northern periphery region of active region AR11360, and the eruption source region was located below a long (about 250 Mm) filament (see Fig. \ref{1}). The eruption started at about 6:00 UT on 29 November 2011, and a slight enhancement accompanied it in the GOES X-ray 1--8 \AA\ flux curve (not shown here). In the eruption source region (before and during the TJ), one can observe a slight emerging negative polarity whose area firstly increased but then followed by a decreasing phase (see the black dashed box in Fig. \ref{2}(a1)-(a6)). In the simultaneous AIA 304 \AA\ images, a loop feature can be recognized around the emerging negative polarity before the TJ (see Fig. \ref{2}(c1) and (c2)). At about 06:00 UT, the TJ had formed around the emerging negative polarity (see Fig. \ref{2}(b6) and (c6)). Refer to online animation of Fig. \ref{2} to better understand the process.

The emergence process is clearly represented by the negative flux measured within the box region as shown in the magnetograms, and the result is plotted in  Fig. \ref{2}(d1). The magnetic flux of the negative polarity reached a maximum value at around 05:18 UT with a growth rate of about $4.18 \times 10^{19} \mathrm{Mx} ~ \mathrm{hr}^{-1}$, after that it started to decrease with a decreasing rate of about $-1.03 \times 10^{19} \mathrm{Mx} ~ \mathrm{hr}^{-1}$. Due to the existence of a large area of positive magnetic field around the negative one, the dynamic evolution of the positive polarity is not evident throughout this period. In analyzing the HMI LOS magnetograms, we note that a positive polarity appeared adjacent to the negative polarity at about 05:19 UT, then it separated from the primary negative polarity and one can identify that it was an another emerging bipole (see Fig. \ref{2}(a4)-(a6)). The appearance of the small bipole was in coincidence with the appearance of a small bright loop-like feature at the same location in the AIA 131 and 304 \AA\ images (see Fig. \ref{2}(b3) and (c3)). During the emerging process of the small bipole, a small jet-like feature can be observed at about 05:53 UT in the AIA 131 and 304 \AA\ images (see Fig. \ref{2}(b4) and (c4)), which might suggest the magnetic reconnection between the loop-like feature and the previous large emerging loop, since hot plasma flow along the large loop was moved from the site of emerging smal-loop (Fig. \ref{2}(c3)) to the south end of the large loop. After this process, the large loop also becomes more visible (see Fig. \ref{2}(c3) versus (c5)). To investigate the correlation between the HMI magnetograms and the response of the EUV images, we analyzed the EUV light curve of the bright loop-like feature and the corresponding magnetic flux changes of the positive polarity. By comparing the curves (see Fig. \ref{2}(d2) and (d3)), we find that both the magnetic flux and the lightcurve measured from AIA 304 \AA\ images underwent a common growth phase and a peak phase after that. In addition, the start (05:26 UT) and peak (05:53 UT) times of the small loop-like feature in AIA 304 \AA\ images were obviously delayed those (05:19 UT and 05:49 UT) of the small positive polarity a few minutes, which suggests that the emerging loop need some time to reach the height of the corona from the photosphere.

\begin{figure}
\centering
\includegraphics[width=8cm]{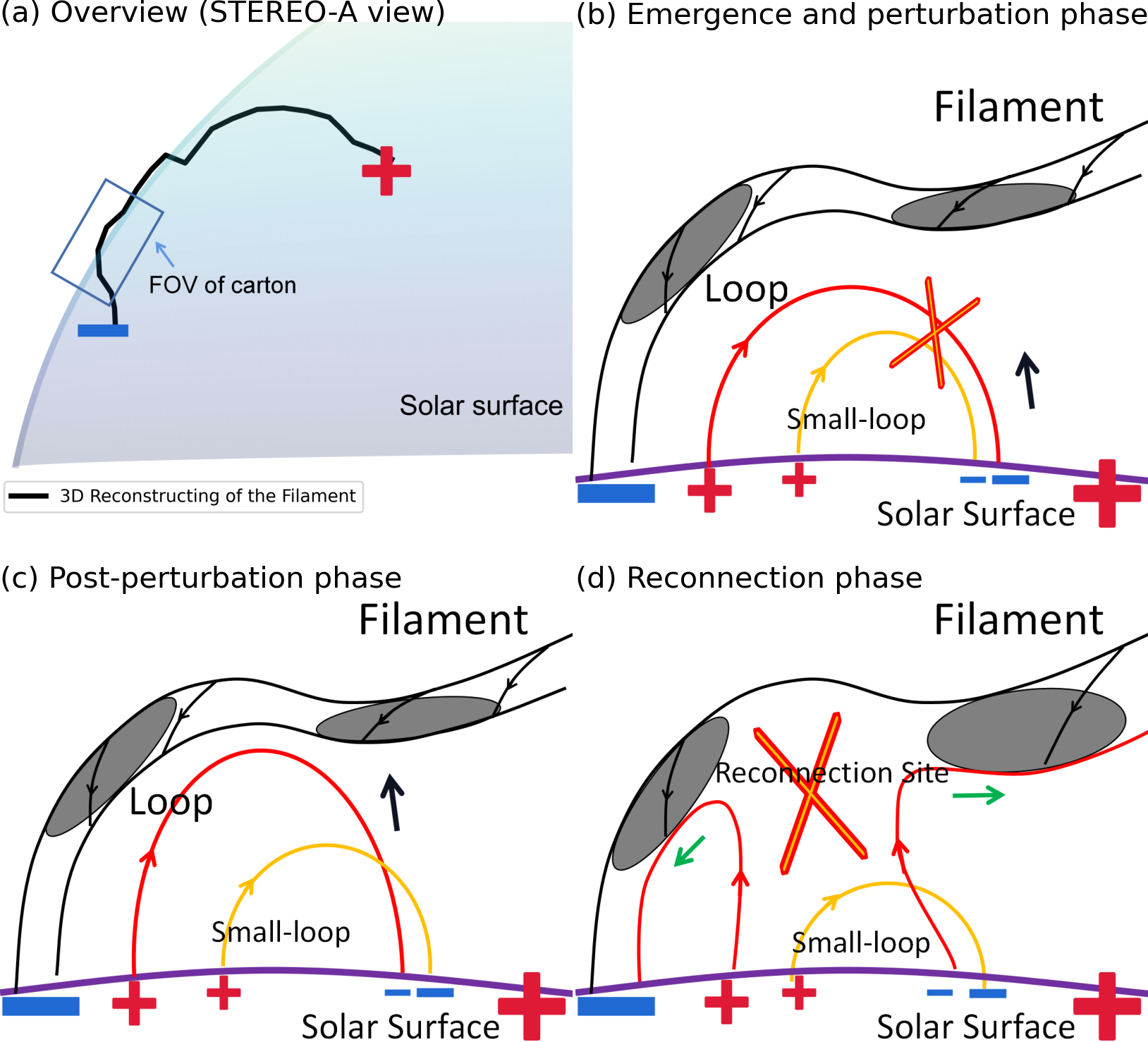}
\caption{ (a) The result of filament 3D reconstruction from STEREO-A view. The blue box indicates the area of carton illustration. (b)-(d) Carton explaining explaining inferred generation process of the TJ. The large red semicircle and small orange semicircle represent the emerging large and small flux loop, respectively. Red X in (b) and (d) represent reconnection site, and red lines in (d) indicate the reconnected field. The green arrows represent plasma flows in both directions along the TJ.
}
\label{3}
\end{figure}

    Fig. \ref{3}(a) shows the 3D reconstruction filament axis using the AIA and EUVI 304 \AA\ image pair at 06:00 UT (black curve), in which the blue box indicates the jet-filament interaction position. Based on our observational results, we propose a cartoon (Fig. \ref{3}(b)-(d)) to illustrate the generation of the TJ. As shown in Fig. \ref{3}(b), a large loop (red) firstly emerges below an overlying large filament. Then, the another small emerging loop (orange) appears below the large one, and reconnection will occur between them (see Fig. \ref{3}(c)). We have also marked the red and orange field lines corresponding to the large and small loops directly on the magnetograms. We can find that on the magnetograms they are at an angle of about 70 degrees, which may be favourable for reconnection to occur (see Fig. \ref{2}(a5-a6)). The appearance of small jet-like feature in the AIA 304 \AA\ images can be viewed as evidence of the reconnection. The newly formed upward-moving reconnected loop (red) further reconnects with the magnetic field of the overlying filament and therefore causes a TJ ejecting along the filament axis (see Fig. \ref{3}(d)).

\subsection{Magnetic structure of the filament-cavity flux rope system}
\label{sec2.2}

\begin{figure}
\centering
\includegraphics[width=8cm]{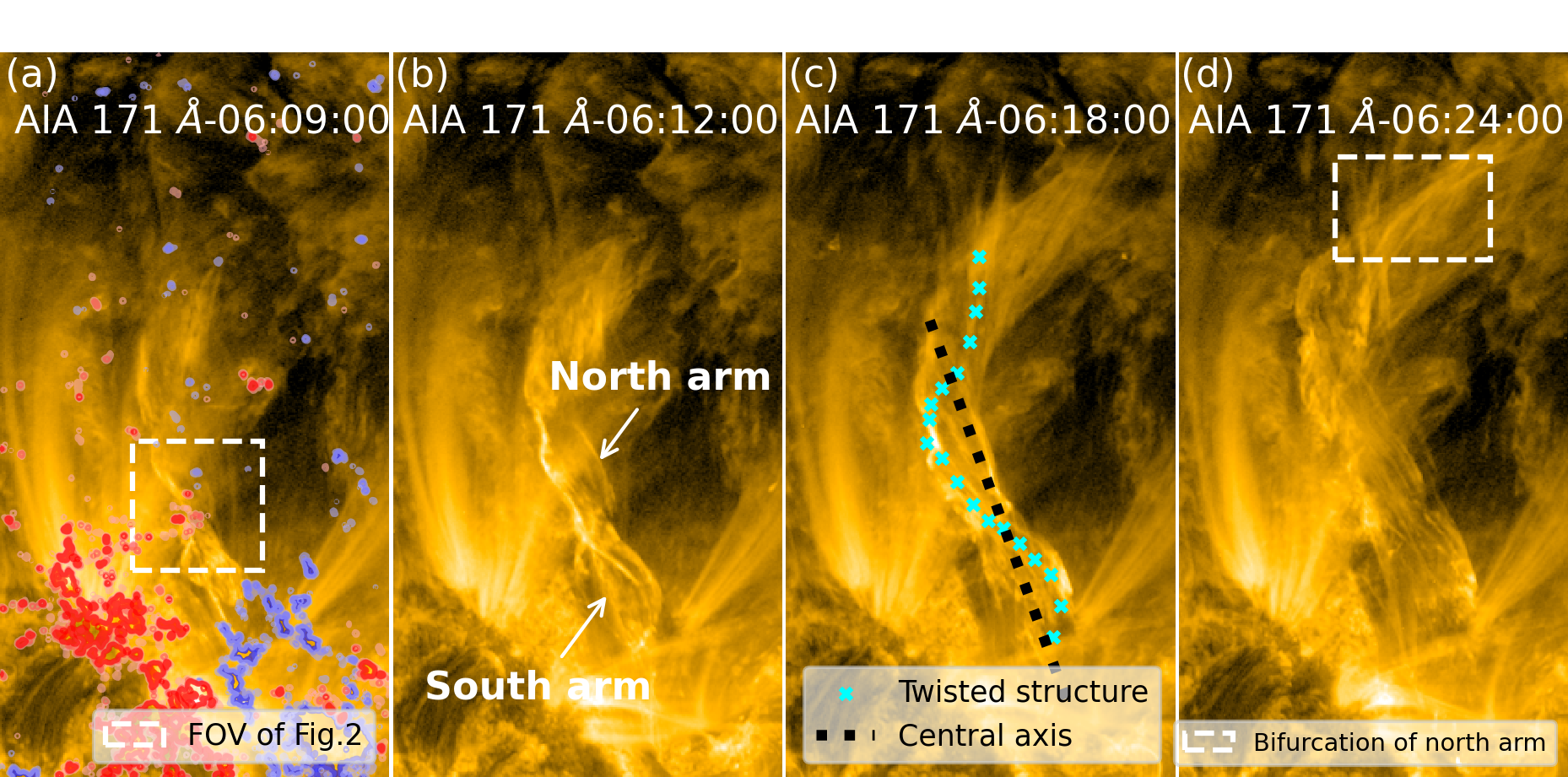}
\caption{Evolution of TJ in filament axis showing twisted structure in AIA 171 \AA\ images  superposed with  HMI LOS magnetogram (only in (a)). The two white arrows in panel (b) indicate the north and south arm of the TJ. The black and cyan dashed lines in panel (c) show the central axis of the filament and the outlined fulx rope structure respectively. The north arm of TJ exhibits a clear bifurcation, which was indicated in the panel (d) by white dashed box. (An animation is provided online.)
}
\label{4}
\end{figure}

\begin{figure}
\centering
\includegraphics[width=8cm]{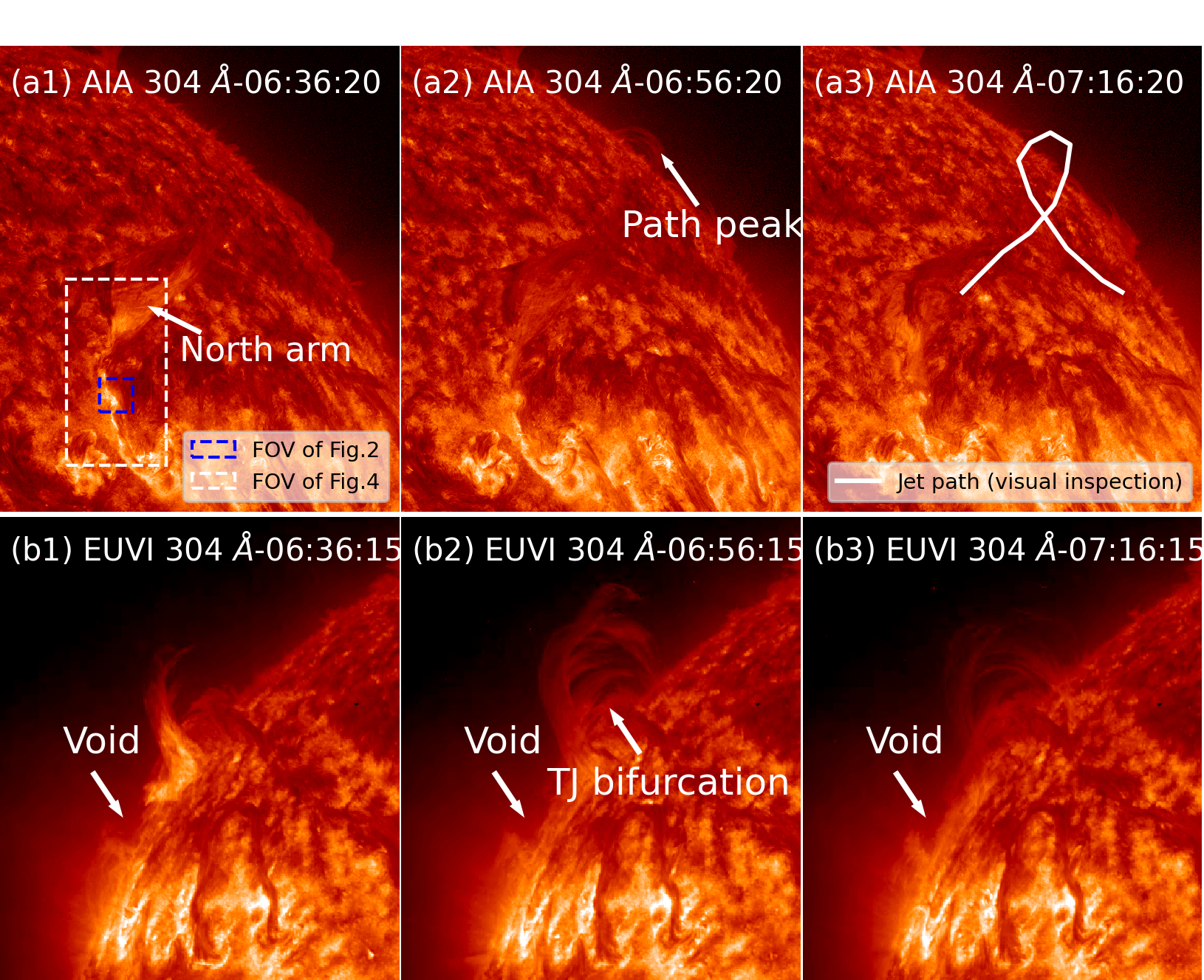}
\caption{Bifurcation of the north arm of the TJ. Panels (a1)-(a3) and (b1)-(b3) are AIA 304 \AA\ time series images from SDO and STEREO view respectively. The blue and white dashed boxes in panel (a1) indicate the FOV of Fig. \ref{2} and Fig. \ref{4}, respectively. In panel (a3) we indicate the propagation path of the jet away from the filament axis with white arrows and marked the peak position it reaches. We also outline the complete jet path with white line in panel (a3). (An animation is provided online.)
} 
\label{5}
\end{figure}

\begin{figure}
\centering
    \includegraphics[width=8cm]{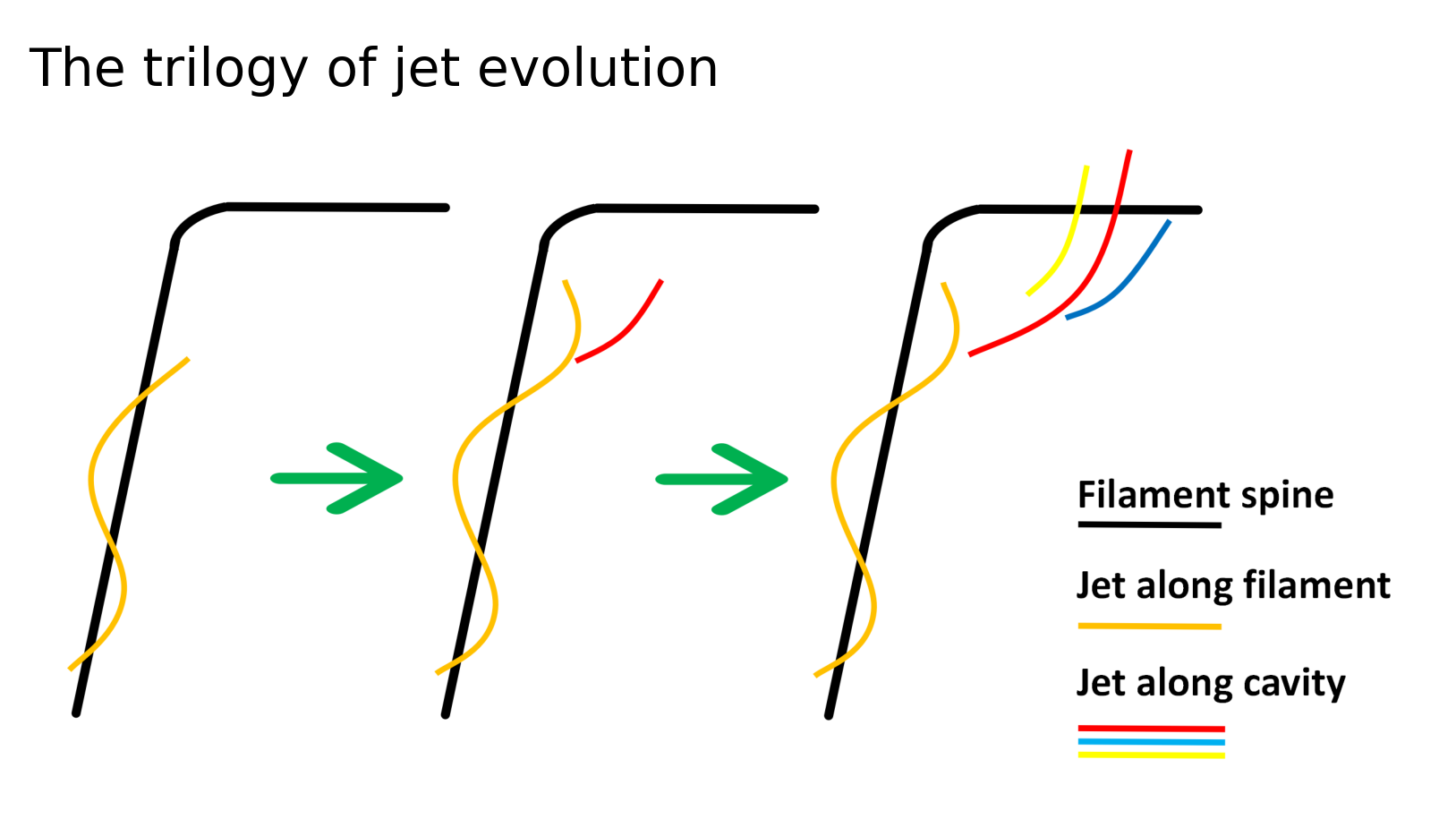}
        \caption{Cartoon diagram explaining the evolution of the jet path. The black line indicates the filament spine. The orange line indicates the jet along the filament, followed by the bifurcation (indicated by the red line). A further bifurcation occurs after the jet enters the coronal cavity, indicated by the yellow, red and blue lines respectively.}
        \label{6}
\end{figure}

\begin{figure*}
\centering
    \includegraphics[width=17cm]{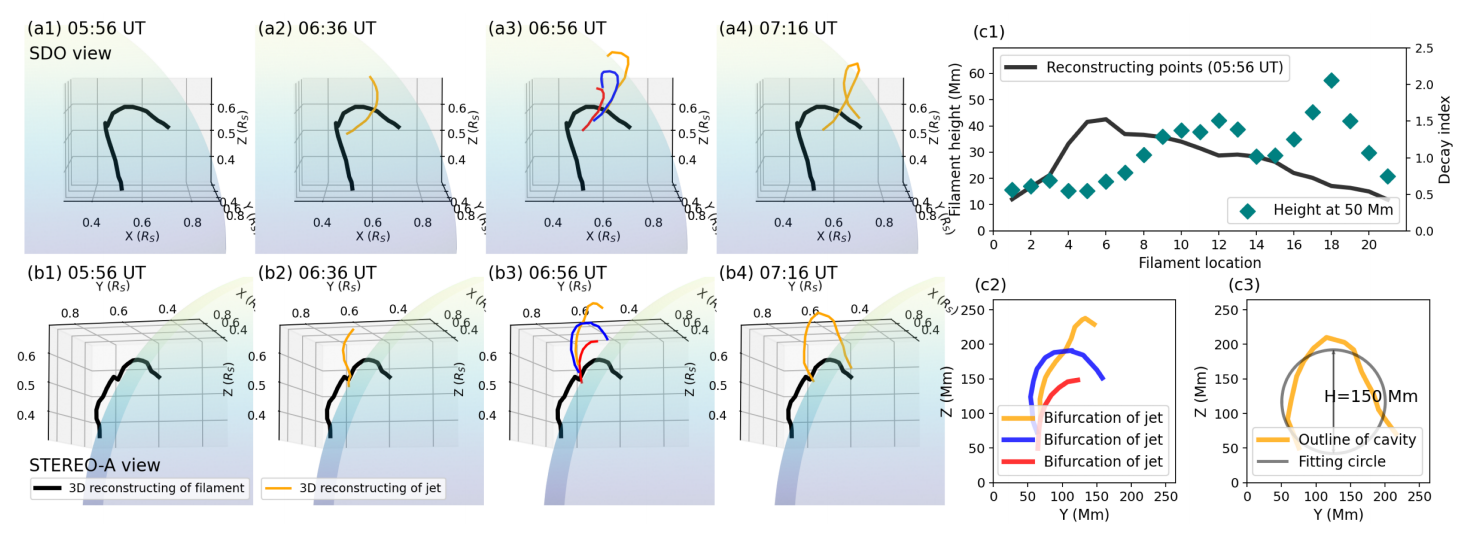}
        \caption{3D reconstruction results and analysis. Panels (a) and (b) are 3D dimensional reconstruction of filament
and jet path from SDO and STEREO view respectively. The black curve in panel (a1) and (b1) indicate the
Undisturbed filament. The subsequent red, blue and yellow curves in panel (a3) and (b3) indicate the different
bifurcated parts of the north arm of TJ. (c1) The variation of the unperturbed filament height (black solid
line) and decay indices (cyan diamonds) derived from the extrapolated potential field along the filament from
south end to north end with selected 21 reconstructed points. On average, the decay index along the filament
is 1.08. (c2) Profile of the Bifurcated north arm of TJ in the YOZ plane obtained by projection. (c3) Complete
projection of the jet trajectory in the YOZ plane and its fitted circle.}
        \label{7}
\end{figure*}

The hot plasma flows accelerated during the loop-filament reconnection exhibited as a TJ along the filament, which also pushed the pre-existing cool filament material to opposite sides from the reconnection site, causing the displacement of the filament material and therefore the formation of a void around the reconnection site as what observed in the STEREO 304 \AA\ images (see Fig. \ref{5}(b1)--(b3)). The void became smaller at about 06:17 UT, which might be caused by the cooling of the hot jet plasma or the retracement of the pushed filament material. To measure the speed of the TJ, a time-space diagram is made along the filament axis (not shown here). It is obtained that the ejecting speeds of the north and south arms were about $92~ \mathrm{km ~s}^{-1}$ and $22~ \mathrm{km~s}^{-1}$, respectively. The significant slower speed of the south arm with respect to the north one was possibly due to the reason that the south part of the filament was too close to the south filament end, where the magnetic and plasma pressures should be very strong since it was rooted in the active region.

An interesting phenomenon observed during the TJ is the rotational motion around the filament axis (see Fig. \ref{4}(b--d) and online animation). Rotational jets are common in the solar atmosphere, from spicule to large scale jets, and the generation of the rotation is generally explained as the transmission of magnetic twist from a closed-loop or flux rope to open coronal loops via magnetic reconnection \citep[e.g.,][]{2011ApJ...735L..43S,2021RSPSA.47700217S}. However, its rotation should be a superficial phenomenon for the present TJ. In our observation, we do not observe obvious twisted structure of the emerging loop. We, therefore, propose that the rotation of the TJ was mainly due to the twisting magnetic structure of the filament, which was highlighted by the moving hot plasma flows. Based on this, we can estimate the twist number of the filament by tracing the TJ mass. As shown by the cyan crosses in Fig. \ref{4}(c), it is easy to count that the magnetic twist of the filament was at least about 3$\pi$ (only considering the part illuminated by the TJ's hot plasma flows).

The north arm of the TJ bifurcated into two parts as it travelled about 50 Mm far away from the reconnection site, and they were moved respectively along the filament axis and the cavity structure (see Fig. \ref{4}(d) and \ref{5}(a1)). With further elevation, the jet material entering the magnetic structure of the coronal cavity bifurcates further and can be clearly distinguished (see the arrows in Fig. \ref{5}(b2)). The part along the cavity peaked at about 06:56 UT and then dropped (see Fig. \ref{5}(a2-a3). The entire TJ path was visually marked in the Fig. \ref{5}(a3). To facilitate the presentation of our understanding of the observed bifurcation, we have made an additional cartoon (see Fig. \ref{6} and legends). The process of jet bifurcation is divided into three steps: first the jets move separately to the north and south, the northward moving jets bifurcate, and the jets entering the magnetic structure of the coronal cavity bifurcate further as they increase in height. This process corresponds to Fig. \ref{4} and \ref{5}, and the reason why the jets bifurcate further is that the magnetic rope structure of cavity in the corona becomes more relaxed with increasing height.

Further, we reconstruct the unperturbed filament and the path of the jet using stereoscopic observations taken by the SDO/AIA and the  STEREO 304 \AA\ paired images, and the panel (a) and (b) of Fig. \ref{7} show the reconstruction results seen from the SDO and STEREO angles, respectively. In Fig. \ref{7} the unperturbed filament axis shows as a black curve, while the path of the jet plasma along the cavity is plotted as red, blue and yellow curves. It should be pointed out that the reconstruction results in Fig. \ref{7}(a3) and (b3) is only qualitatively analyzed,  as the AIA 304 \AA\ image does not identify the bifurcation very well. It can be seen in Fig. \ref{7}(a4) that the cavity of the filament had a twisted structure, and its twist was at least one turn. Therefore, the total twist of the entire filament should be greater than 5$\pi$ when we directly assume that the twist of the cavity is approximately equal to the right part of filament (not illuminated by the TJ's hot plasma flows). Further, we can find that the outer coronal cavity is less twisted relative to the inner filament, which suggests that the twist is non-uniform and decreasing with distance from the filament
axis.

We plot the real height of the entire filament based on our reconstruction result in Fig. \ref{7}(c1). It can be seen that the heights of the two ends were about 10 Mm above the solar surface, and the highest part was about 40 Mm. In addition, the height distribution of the filament was asymmetric with respect to the middle point. The highest part was closer to the south end, which is rooted in the active region. We further analyze the complete trajectory of the TJ's north arm at 07:16 UT. Since this part of the filament is almost parallel to the X-axis of the 3D coordinate system (see Fig. \ref{7}(b4)), we project the jet trajectory directly into the YOZ plane; so that we obtain the coronal cavity profile as seen from the limb. We then fit the cavity and obtain that the diameter of the cavity is about 150 Mm (see Fig. \ref{7}(c3)), which is very close to the height (168 Mm) of the coronal cavity observed a few days later at the limb in AIA 193 \AA\ images.

\section{Discussions}
The bifurcation of solar jets has been reported in several previous studies \citep[e.g.,][]{1999SoPh..190..167A,2012ApJ...745..164S,2022ApJ...926L..39D}, and such a phenomenon has several different explanations, for example, dynamic response to the rapid transport of twist along magnetic field \citep{1996ApJ...464.1016C}, the development of instability such as Kelvin-Helmholtz Instability \citep{1999SoPh..190..167A}, and the uncoupling of the erupting filament's two legs in filament-driven jets \citep{2012ApJ...745..164S}. In the present observation, we propose that the bifurcation of the TJ's north arm can be explained by the magnetic flux rope model of the filament, where the filament and the hosting cavity are considered as two interconnecting elements of one flux rope system \citep{2006JGRA..11112103G}. As the TJ occurred close to the south end of the filament, the magnetic structures of the filament and the cavity are not yet been spatially separated; therefore, the plasma along different magnetic field lines looks like along a single magnetic structure. However, when the jet reaches the middle of the filament where the cavity and the filament components are well spatially separated, one can clearly distinguish the plasma flows along different magnetic structures and therefore formed the observed bifurcation effect of the jet. Thus, the bifurcation of solar jets can also be interpreted as plasma flows along different magnetic field lines of the same magnetic system, such as a filament-cavity flux rope. Vice versa, the bifurcation of solar jets can also be used to diagnose the magnetic structure of coronal structures.

The magnetic twist of the filament was at least 5$\pi$, which has excessed the kink instability threshold (2.5$\pi$) for a line-tying configuration. However, the filament kept stable despite the disturbance rose by the TJ. This discrepancy can be reconciled by considering the physical properties of the filament itself and the surrounding coronal condition. As suggested by \citep{2005ApJ...630L..97T}, for overlying coronal magnetic fields decreased gradually with respect to the height (i.e., small decay indexes), filaments beneath them tend to be more stable than that confinement field rapidly decreased in height. Fig. \ref{7}(c1) also shows the decay indices at the height of 50 Mm derived from the extrapolated potential field along the filament from south end to north end. The average decay index of the overlying coronal magnetic field of the filament is 1.08, which is far less than the threshold (1.5) of torus instability \citep{2006PhRvL..96y5002K}. Especially at the south end of filament where TJ ejects, the decay indexes are all less than 1. Therefore, we believe that the small decay indexes play an essential role in maintaining the stability of filaments \citep{2008ApJ...679L.151L,2021ApJ...923...45Z}. The height is another reason for the stability of filaments. For example, statistical results suggest that filaments with a longer length and lower height are more stable than those with shorter length and higher height \citep[e.g.,][]{2012ApJ...744..168L}. In the present observations, both the length (250 Mm) and height (40 Mm) parameters suggest that the current filament should be a stable one. Recent statistical work on small samples has also shown that long quiescent filaments tend to carry more twist than short active-region filaments \citep{2021ApJ...917...81G}. Those may be the reasons why the filament with high twist did not erupt, although it was disturbed by the TJ.

\section{conclusions}
In this letter, we present the first stereoscopic observation of a TJ triggered by an emerging loop underneath an overlying large filament. It is found that the TJ was caused by the magnetic reconnection between the emerging loop and the overlying filament. By tracing the ejecting plasma flows of the TJ inside the filament, we not only measured that the magnetic twist stored in the filament was at least 5$\pi$, but also found that the fine magnetic structure of the filament-cavity flux rope system is in well agreement with the theoretical results of MFR models. The high twisted but stable filament can be understood by considering its longer length and lower height properties, and the small average decay indexes of the overlying coronal field. In addition, the flux rope structure revealed in the present study is also consistent with some statistical results (\cite{2017ApJ...835...94O}). Our observation provides a new perspective for understanding MFRs and a new method for diagnosing magnetic field structures of on-disk filaments.

\section*{Acknowledgements}

The authors would like to acknowledge the SDO, GONG, GOES, and STEREO science teams for providing the data. This work is supported by the Natural Science Foundation of China (11922307, 12173083, 11773068, 11633008), the Yunnan Science Foundation for Distinguished Young Scholars (202101AV070004), the National Key R\&D Program of China (2019YFA0405000),  the Specialized Research Fund for State Key Laboratories, the West Light Foundation of Chinese Academy of Sciences, and the Yunnan Key Laboratory of Solar Physics and Space Exploration (202205AG07009).

\section*{Data availability}

This research used the SunPy \citep{sunpy_community2020}software package to present the results of our observation. All data is public and available through SunPy's download interface.
%%%%%%%%%%%%%%%%%%%%%%%%%%%%%%%%%%%%%%%%%%%%%%%%%%



%\bibliographystyle{mnras}
%\bibliography{ref} % if your bibtex file is called example.bib

%\begin{thebibliography}{99}
%\end{thebibliography}

%%%%%%%%%%%%%%%%%%%%%%%%%%%%%%%%%%%%%%%%%%%%%%%%%%

%%%%%%%%%%%%%%%%% APPENDICES %%%%%%%%%%%%%%%%%%%%%

%%%%%%%%%%%%%%%%%%%%%%%%%%%%%%%%%%%%%%%%%%%%%%%%%%

% Don't change these lines
\bsp	% typesetting comment
\label{lastpage}
\end{document}